\documentclass{article}

\PassOptionsToPackage{numbers, compress}{natbib}
\usepackage[preprint]{neurips_data_2023}

\usepackage[utf8]{inputenc} % allow utf-8 input
\usepackage[T1]{fontenc}    % use 8-bit T1 fonts
\usepackage{hyperref}       % hyperlinks
\usepackage{url}            % simple URL typesetting
\usepackage{booktabs}       % professional-quality tables
\usepackage{amsfonts}       % blackboard math symbols
\usepackage{nicefrac}       % compact symbols for 1/2, etc.
\usepackage{microtype}      % microtypography
\usepackage{xcolor}         % colors
\usepackage{graphicx}
\usepackage{amsmath}
\usepackage{amsfonts}
\usepackage{booktabs}
\usepackage{caption}
\usepackage{subcaption}
\usepackage{adjustbox}

\title{Smiles2Dock: an open large-scale multi-task dataset for ML-based molecular docking}

\author{%
  Thomas Le Menestrel \\
  Department of Computational Engineering \\
  Stanford University\\
  Stanford, CA 94305 \\
  \texttt{tlmenest@stanford.edu} \\
  \And
  Manuel A. Rivas \\
  Department of Biomedical Data Science \\
  Stanford University\\
  Stanford, CA 94305 \\
  \texttt{mrivas@stanford.edu} \\
}

\begin{document}

\maketitle

\begin{abstract}
\label{sec:abstract}

  Docking is a crucial component in drug discovery aimed at predicting the binding conformation and affinity between small molecules and target proteins. ML-based docking has recently emerged as a prominent approach, outpacing traditional methods like DOCK and AutoDock Vina in handling the growing scale and complexity of molecular libraries. However, the availability of comprehensive and user-friendly datasets for training and benchmarking ML-based docking algorithms remains limited. We introduce Smiles2Dock, an open large-scale multi-task dataset for molecular docking. We created a framework combining P2Rank and AutoDock Vina to dock 1.7 million ligands from the ChEMBL database against 15 AlphaFold proteins, giving us more than 25 million protein-ligand binding scores. The dataset leverages a wide range of high-accuracy AlphaFold protein models, encompasses a diverse set of biologically relevant compounds and enables researchers to benchmark all major approaches for ML-based docking such as Graph, Transformer and CNN-based methods. We also introduce a novel Transformer-based architecture for docking scores prediction and set it as an initial benchmark for our dataset. Our dataset (\url{https://huggingface.co/datasets/tlemenestrel/Smiles2Dock}) and code (\url{https://github.com/rivas-lab/Smiles2Dock}) are publicly available to support the development of novel ML-based methods for molecular docking to advance scientific research in this field.
\end{abstract}

\section{Introduction}

\subsection{Overview of docking}

Molecular docking is a computational method used to predict the non-covalent binding of a small molecule, or ligand, to a biological macromolecule, typically a protein, which is crucial for understanding biochemical processes and designing drugs \cite{meng2011molecular}. It predicts the optimal orientation and position of a ligand within a protein's binding site, where these interactions can stabilize the complex, thereby inhibiting or enhancing the protein's function. This process mimics the natural molecular recognition and binding phenomena that occur in biological systems \cite{morris2008molecular}. 

Docking is crucial for drug development as it allows researchers to dramatically reduce the number of candidate molecules that need to be synthesized and tested, by screening through large libraries of compounds and identifying those most likely to bind to the target protein \cite{gschwend1996molecular, Stanzione2021}. Moreover, effective docking results not only forecast where and how well a ligand binds, but also provide insights into the nature of the binding affinity and specificity. By simulating these interactions, scientists can infer the biological activity of new compounds, further accelerating the drug development process \cite{meng2011molecular}.

Docking algorithms output scores, typically in kcal/mol, indicating the predicted binding affinity between the ligand and protein, with more negative values suggesting stronger binding. The algorithm also provides ranked binding poses, showing the spatial arrangements of the ligand within the receptor's active site. These outputs help identify potential lead compounds by highlighting those with the most favorable predicted interactions.

\subsection{Chemistry concepts and terminology}

\textbf{Proteins and ligands:} Proteins are large, complex molecules that play critical roles in the body, ranging from catalyzing metabolic reactions to providing structural support in cells. Ligands are smaller molecules that can bind to proteins, influencing their function and activity, often used as drugs to modulate protein behavior in therapeutic treatments.

\textbf{Binding sites:} A binding site is a specific location on a protein where a ligand can form a stable interaction. This site is typically a pocket or groove on the protein’s surface that is complementary in shape and chemical properties to the ligand, allowing specific and selective binding. The interactions between the protein and ligand at the binding site involves various forces including hydrogen bonds, hydrophobic forces, and van der Waals interactions. The specificity and strength of these interactions are crucial for the biological function of the protein, as they can significantly influence the protein's activity, stability, and regulation. Understanding the structure and properties of binding sites is essential for drug design and protein engineering, as modifications to the binding site can alter how effectively a ligand can interact, leading to changes in the therapeutic efficacy or biological response of the protein \cite{Jiang2019}. This knowledge is used in molecular docking studies to predict how and where ligands might interact with proteins.

\textbf{Conformational space:} The conformational space refers to the range of different arrangements that a molecule can adopt in three-dimensional space due to the rotation around its single bonds. This flexibility allows molecules, especially complex organic compounds, to exist in multiple conformations, each differing in spatial orientation but not in the connectivity of atoms. The exploration of conformational space is crucial in understanding how a molecule behaves in different environments, how it interacts with other molecules, and how it fits into enzyme or receptor sites in biological systems. In drug design, assessing the conformational space of a molecule helps predict its binding affinity and pharmacokinetic properties, as different conformations can have different biological activities. Vast conformational spaces pose significant challenges in computational chemistry, as algorithms strive to efficiently sample as many feasible conformations as possible to accurately predict a molecule’s behavior in a biological context \cite{Shin2011}.

\subsection{ML-based docking over traditional methods}

As the scale of molecular libraries expands dramatically in drug discovery, the need for faster and more efficient docking tools has become evident. Traditional software like DOCK, which relies on a geometric matching algorithm to fit ligands into protein binding sites, AutoDock Vina, known for its use of gradient optimization methods to predict binding poses, and Glide, which performs a systematic search of the conformational, orientational, and positional space of the docked ligand, have proven too slow for handling modern large-scale libraries \cite{Coleman2013, Trott2009, friesner2004glide}. In response, researchers are increasingly turning towards Machine Learning (ML) based approaches to enhance docking efficiency. Machine learning models can learn from vast datasets of molecular interactions, enabling them to predict binding outcomes much faster than traditional methods, achieving speedups of 10 to 100 times \cite{Crampon2022}. 

Several ML approaches have been tried. The most prominent one is Graph Neural Networks (GNNs), which directly model the molecular structure of proteins and ligands as graphs where atoms are nodes and bonds are edges \cite{Jiang2022, wang2021protein, han2021quality}. 

An extension of GNNs is Graph Convolutional Networks (GCNs), which apply convolutional neural network concepts to graph-structured data, allowing the network to capture the topological features of molecules and their potential interactions with proteins \cite{torng2019graph}. While originally developed for natural language processing, Transformers have been adapted to handle molecular data by treating atoms or fragments as sequence elements. These models can be pretrained on large sets of SMILES strings and, capture long-range interactions within molecules and between molecules and proteins and finaly represent proteins and ligands as embedding matrices or vectors \cite{guo2022vitrmse, guo2023vitscore, chu2024flexible}. Another solution is to use Computer Vision models such as 3D Convolutional Neural Networks (3D CNNs). They extend the idea of convolutional networks into three dimensions, which is natural for interpreting the spatial structure of molecules and their interactions within the 3D space of protein binding sites \cite{zheng2019onionnet, mcnutt2021gnina, jiang2020guiding}. Reinforcement learning approaches have also been tested by leveraging the asynchronous advantage actor-critic (A3C) model, a novel reinforcement learning approach to enhance protein-ligand docking that employs an actor model to guide search actions and a critic model to evaluate these actions, significantly improving the accuracy of binding site and docking predictions \cite{chong2021reinforcement, aderinwale2022rl}.

\section{Related work}

The downside of ML-based methods is the amount of data required to train those. To remedy this, several groups have attempted to build open-source docking datasets by using docking software predictions as inputs for ML models \cite{Clyde2023, GarcaOrtegn2022, Luttens2023}. However, available large-scale docking datasets have several limitations, notably scale, ease of use and lack of generalizability. 

Some focused on a specific set of proteins linked to a certain disease (e.g. SARS-COV2 proteome), greatly reducing generalization capabilities for ML models trained on those. Others used a number of ligands not in the scale of modern compound libraries, which often have millions of data points, and did not use well-known extensively tested chemical libraries. Finally, certain datasets are it difficult to be used by all different ML-based docking methods, for example by not using the SMILES format to represent molecules (directly formatted as .sdf or .pdb in 3D) \cite{mysinger2012directory}. This prevents the usage of Transformer models, which take as input a SMILES string and turn it into an embedding vector or matrix. A summary of the limitations and strengths of 3 well-known molecular docking datasets alongside Smiles2Dock is available in Table 1.

\begin{table}[h]
    \caption{Summary of datasets used for molecular docking}
    \begin{tabular}{lllll}
        \hline
        \textbf{Name} & \textbf{Size} & \textbf{Number of ligands} & \textbf{Number of proteins} & \textbf{SMILES format} \\
        \hline
        Smiles2Dock & \textbf{26M} & \textbf{1.7M} & 15 & Yes \\
        \hline
        DOCKSTRING & 15M & 260k & 58 & Yes \\
        \hline
        LIT-PCBA & 3M & 200k & 15 & Yes \\
        \hline
        DUD-E & 233k & 23K & \textbf{102} & No \\
        \hline
    \end{tabular}
    \label{tab:docking_datasets}
\end{table}

To address these limitations, we docked a set of 15 proteins from AlphaDock with 1.7M molecules from the CHeMBL database. We made our P2Rank binding site predictions available for each protein to allow researchers to reuse them for docking. In our work, we represent the molecules from CHeMBL in the SMILES format. The SMILES (Simplified Molecular Input Line Entry System) format is a notation method that encodes the structure of a molecule using a simple line of text, which describes the arrangement of atoms and bonds within the molecule. This compact and readable format allows for easy input, storage, and exchange of information across various chemical informatics software and databases. By using this format, Smiles2Dock can accommodate Transformer-based methods, which are pretrained on SMILES string and generate embeddings from them to represent molecules. Using SMILES strings for our dataset also allows us to benchmark Computer Vision and graph-based methods, as chemists can use tools such as OpenBabel \cite{OBoyle2011} and RDKit \cite{RDKit} to generate a 3D representation of the target ligand and pass that as input to their models. Finally, we make Smiles2Dock available on HuggingFace, allowing researchers to load our dataset in a Jupyter Notebook using only two lines of code and a single Python library \cite{lhoest2021datasets, kluyver2016jupyter}. 

\section{Dataset construction}

\subsection{Protein and ligand datasets}

\textbf{Alphafold:} AlphaFold is an ML system developed by DeepMind designed to predict protein structures and solve the protein folding problem, which involves determining a protein's three-dimensional shape from its amino acid sequence \cite{jumper2021highly}. In the 14th Critical Assessment of Structure Prediction (CASP14) held in 2020, AlphaFold achieved a median Global Distance Test (GDT) score of 92.4 out of 100, significantly outperforming other methods and marking a substantial leap in predictive accuracy. Its predictions have been extensively validated, with a reported root-mean-square deviation (RMSD) of around 1.5 Ångströms for many proteins, comparable to experimental methods like X-ray crystallography and cryo-electron microscopy, while being significantly cheaper and faster. 

AlphaFold's approach leverages neural networks trained on known protein structures to predict the spatial relationships between amino acids, effectively modeling how a protein folds in nature and providing reliable protein structure predictions, which are essential for understanding biological processes and developing new therapeutics. AlphaFold's predictions have been validated through rigorous testing and are particularly valuable for creating molecular docking datasets, as they provide detailed structural information crucial for predicting protein-ligand interactions. 

\textbf{ChEMBL:} ChEMBL is a bioactivity database maintained by the European Bioinformatics Institute, containing detailed information on the biological activity of 2.3M small molecules \cite{gaulton2012chembl}. It is widely used for drug discovery and development, offering data on compound properties, target interactions, and pharmacological profiles. ChEMBL's extensive and curated datasets are invaluable for creating molecular docking datasets, as they provide a rich source of known interactions and compound structures. 

It encompasses over 2.3 million bioactive molecules, providing detailed chemical structures, calculated properties (such as logP, molecular weight, and Lipinski's Rule of Five parameters), bioactivity data (including binding constants, pharmacological effects, and ADMET profiles), and annotations on target interactions, clinical progress, and therapeutic indications. Using molecules from ChEMBL facilitates the validation and training of docking algorithms, enhances predictive accuracy, and supports the identification of potential drug candidates by leveraging robust bioactivity data.

\subsection{Framework outline}

\begin{figure}[ht]
\centering
\includegraphics[width=1\textwidth]{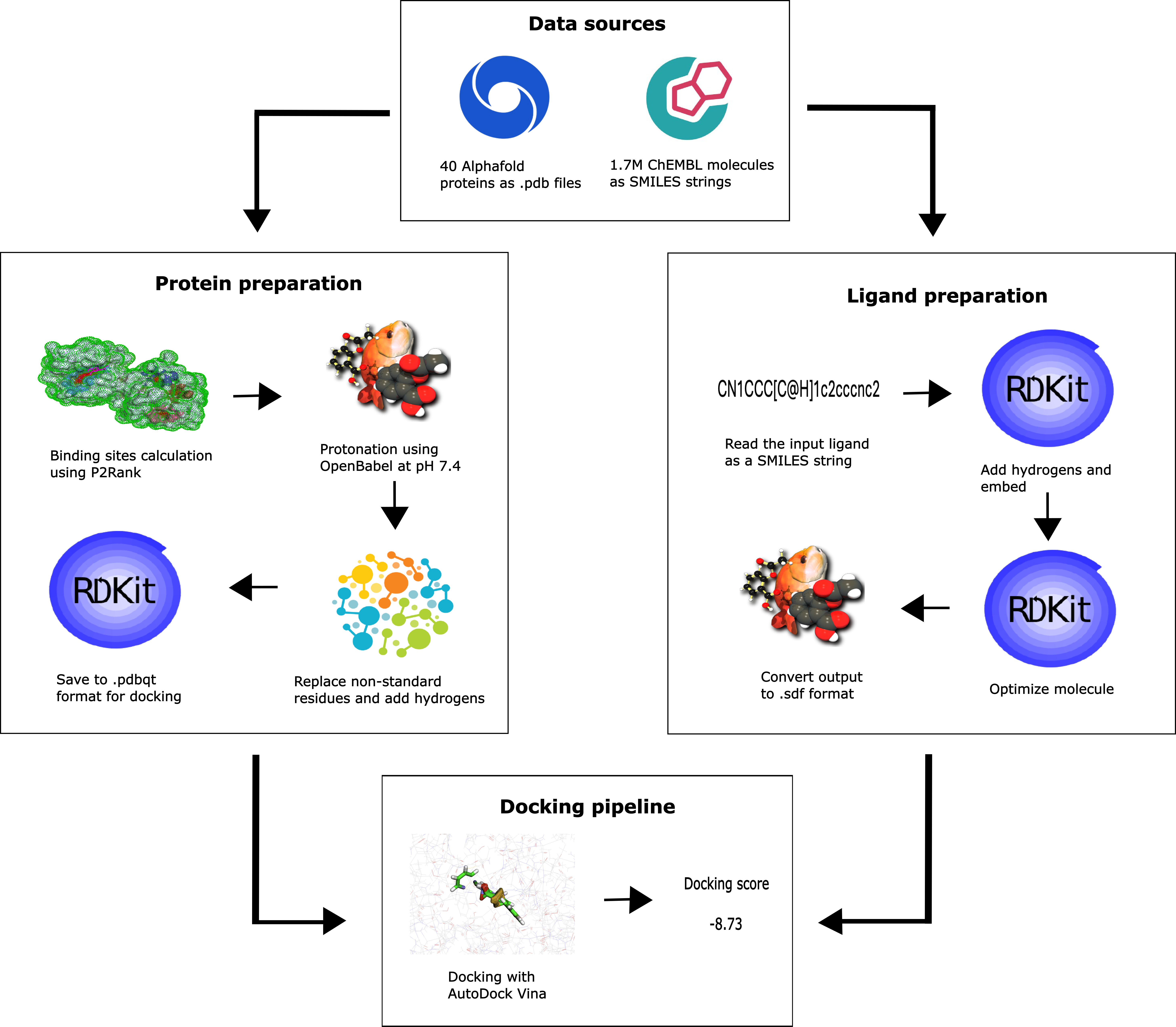}
\caption{{{\bf Diagram of the methodology followed for this project.}}}
\label{fig:interactions}
\end{figure}

\textbf{P2Rank:} P2Rank is an ML model for predicting ligand-binding sites on proteins by analyzing surface patches based on features like hydrophobicity, electrostatic potential, and geometric arrangement of atoms \cite{Krivk2018}. Each protein surface is segmented into patches, with the random forest model assessing the likelihood of each patch being a binding site based on the extracted features. This method allows for efficient and rapid identification of potential binding sites, making P2Rank particularly useful in drug discovery and molecular docking studies.

\textbf{AutoDock Vina:} AutoDock Vina is a popular molecular docking software widely used in computational chemistry for predicting the interactions between a protein and a ligand. It uses a scoring function to estimate the strength and stability of a ligand when docked into a protein’s binding site. One of the key features of AutoDock Vina is its use of an iterative search algorithm that efficiently explores possible docking poses (orientations and conformations of the ligand within the binding site). The "exhaustiveness" parameter in Vina plays a crucial role in this process, as it controls how thoroughly the search space is explored. A higher exhaustiveness value increases the probability of discovering the best possible docking pose, at the cost of increased computational time and resources, effectively allowing researchers to balance accuracy and computational efficiency. 

\subsection{Methodology}

In our study, we developed a dataset of molecular docking scores using a comprehensive framework to ensure precise predictions of protein-ligand interactions. The framework can be seen on Figure 1. First, we selected a set of 30 proteins from the AlphaFold database based on their identification as therapeutic targets in genetic association studies \cite{tanigawa2019components, sakaue2021cross, tanigawa2020rare, deboever2018medical}. This highlights their potential relevance in disease mechanisms and therapeutic intervention. We then used P2Rank to predict their binding sites and only kept proteins for which we had at least one site with a probability above 50\%. The remaining set of 15 proteins was then refined by replacing nonstandard residues, removing heterogens and adding hydrogens at physiological pH using OpenBabel, RDKit and OpenMM \cite{eastman2017openmm}.

For ligands, we downloaded the CHeMBL database and used the 2.3M SMILES strings available. Out of this set, around 20\% could not be processed by AutodockVina because of errors, either when converting SMILES strings to .sdf files using RDKit or due to atom types incompatible with Autodock. This left us with a set of around 1.7M ligands to dock. Those were then optimized through geometric adjustments and format conversions using RDKit and OpenBabel.

Lastly, we used AutoDock Vina through its Python extension to perform the docking, specifying 5 poses per ligand and an exhaustiveness level of 8 \cite{Eberhardt2021}. The computations were executed on a High-Performance Computing (HPC) cluster, taking approximately 45 days to complete and 600,000 CPU hours. Finally, we extracted and compiled all the docking scores to construct our dataset. 

\subsection{Dataset splits}

To allow for training and tuning ML models, we decided to split our dataset into 3 folds using a traditional 70-10-20 split (70 \% for training, 10 \% for validation and 20 \% for testing). This allows for hyperparameter tuning on the validation set and then final testing on the remaining 20 \%. The exact numbers can be found in Table 2.

\begin{table}[ht]
      \centering
    \caption{Distribution of docking scores.}
    \begin{tabular}{lrrr}
      \toprule
      \label{sample-table}
      \centering
     & Train & Validation & Test \\
    \midrule
    Very Strong & 2188 & 312 & 637 \\
    Strong & 218316 & 31424 & 62404 \\
    Medium+ & 9782630 & 1397159 & 2795557 \\
    Medium- & 7931410 & 1133130 & 2266133 \\
    Weak & 215727 & 30904 & 61538 \\
    Very Weak & 187330 & 26729 & 53046 \\
    \bottomrule
    Total & 18337601 & 2619658 & 5239315\\
    \end{tabular}
\end{table}

The definition of a "strong" or "weak" docking scores varies greatly depending on the type of protein and the experimental setup. To provide a unified view of our dataset, we classified scores using the overall distribution for each protein. We computed the mean and standard deviation for each protein and classified the scores in 6 categories. Medium+ and Medium- are for scores within 1 standard deviation of the mean, while Strong and Very strong are for scores below 2 and 3 standard deviations of the mean respectively and Weak and Very Weak are for scores above 2 and 3 standard deviations of the mean (the more negative the docking score, the stronger the binding).

\subsection{Distribution of docking scores}

We investigated further the distribution of the scores. Our initial hypothesis after looking at Figure 2 was that scores for each protein were normally distributed. We performed Shapiro-Wilk tests to test for normality of scores distribution for each protein but all p-values were below the 0.05 significance threshold. \cite{razali2011power}. We discovered, by computing a Q-Q plot (Figure 3), that the distribution was heavily right-skewed, which was also confirmed by computing the skewness of the distribution of scores for each protein, with values ranging from 5 to 20 (heavily right skewed) \cite{marden2004positions}. This can also be seen by looking at the number of scores categorized Strong and Very Strong in Table 2 versus Weak and Very Weak, where the ratio of Very Strong to Very Weak scores in the Train set is about 86 to 1. Finally, we tested for right-skewed distributions by performing a Kolmogorov-Smirnov test for goodness of fit using Log-Normal and Weibull distributions but again found p-values for all proteins below the significance threshold required \cite{berger2014kolmogorov}.

\begin{figure}[h]
\centering
\begin{minipage}[b]{0.45\textwidth}
    \centering
    \includegraphics[width=\textwidth]{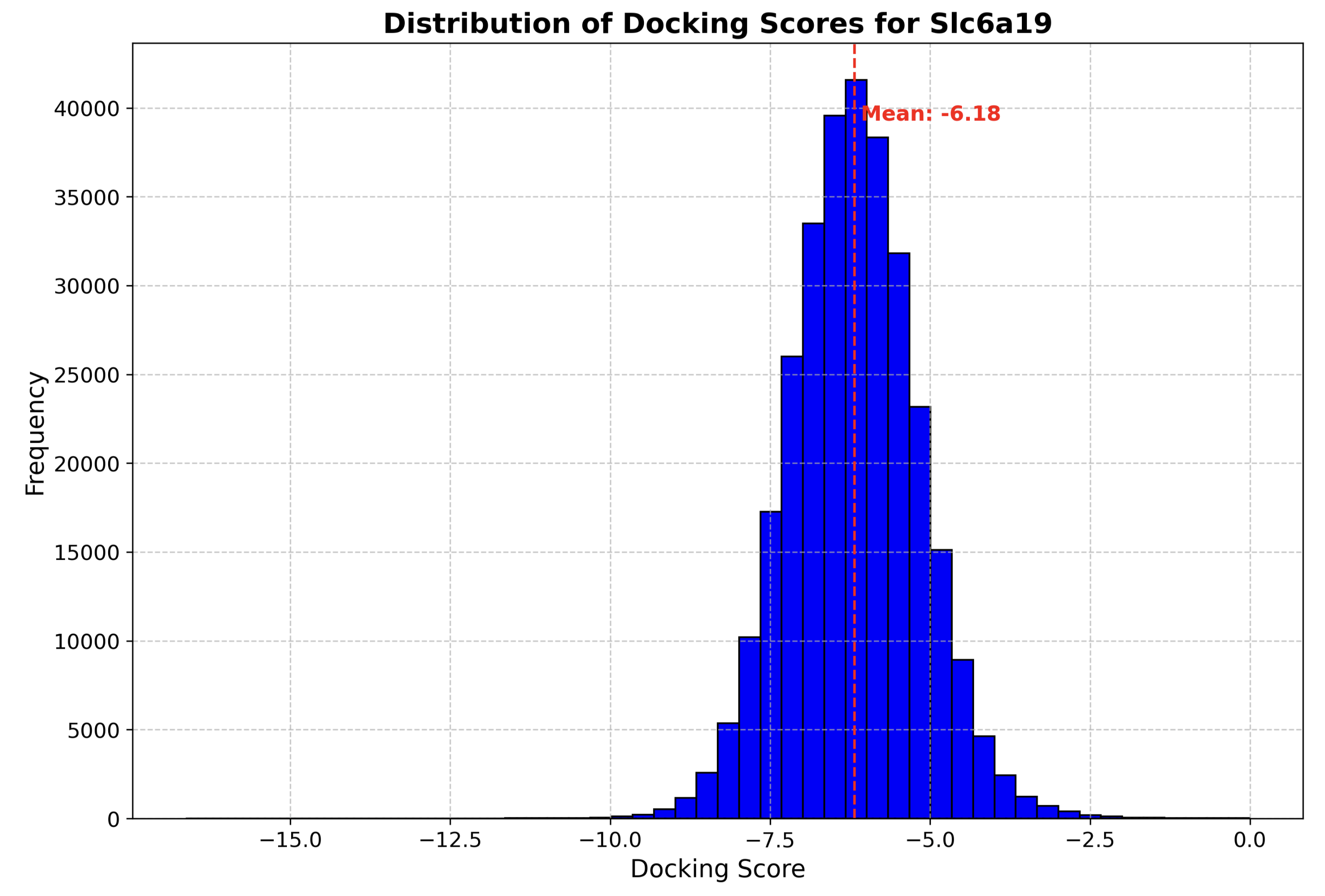}
    \caption{Distribution of docking scores for protein Slc6a19.}
    \label{fig:slc6a19}
\end{minipage}
\hfill
\begin{minipage}[b]{0.45\textwidth}
    \centering
    \includegraphics[width=\textwidth]{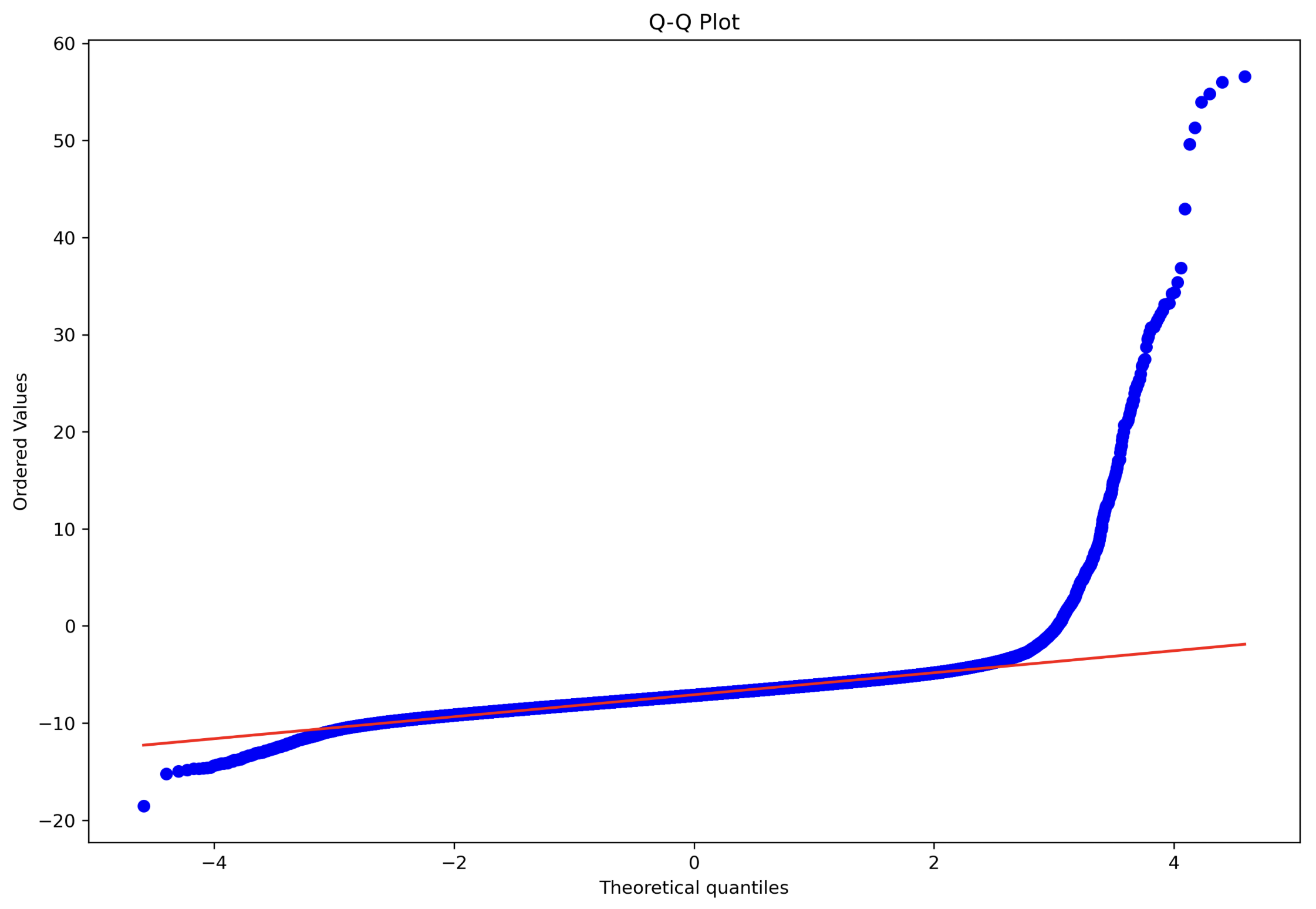}
    \caption{QQ plot of docking scores for protein Slc6a19.}
    \label{fig:qqplot}
\end{minipage}
\end{figure}

\section{Experiments}

\subsection{Transformer architecture}

To inform ML researchers and benchmark our dataset, we built a novel Transformer method to predict docking scores. We took an embedding-based approach and used two foundation models to encode the protein and the ligand and perform the docking (Figure 4).

\begin{figure}[ht]
\centering
\includegraphics[width=0.9\textwidth]{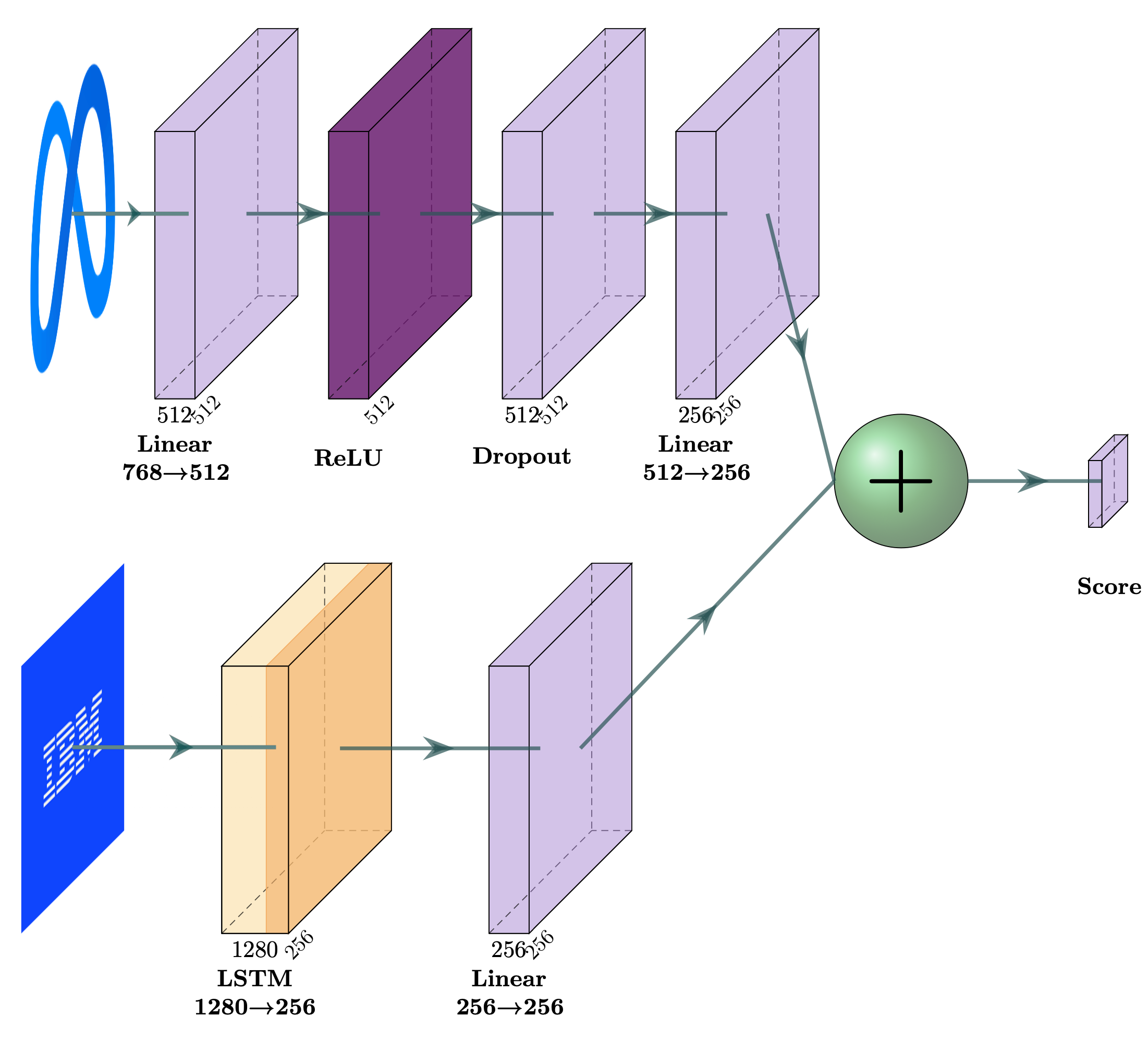}
\caption{{ Architecture of the hybrid LSTM-FFN protein-ligand model.}}
\label{fig:interactions}
\end{figure}

\subsection{Foundation models}

The first one is ESM2, developed by Facebook AI Research (FAIR), which was developed using a pretraining framework involving 86 billion amino acids across 250 million protein sequences \cite{Lin2022}. The representation space developed is organized across multiple scales, capturing details from the biochemical properties of amino acids to the distant homology between proteins. It encodes information about both secondary and tertiary structures, which can be discerned through linear projections. The model uses a masked language model to learn the properties of amino acids and a community propagation-based clustering algorithm to connect these properties to the protein's overall structure and function. ESM2 can make accurate predictions for over 40,000 protein isoforms and is used in tasks like MHC peptide binding prediction and single-sequence protein structure prediction. This representation learning generates features that are versatile across various applications, facilitating the supervised predictions of mutational effects and secondary structures, as well as enhancing the best existing features for predicting long-range contacts.

The second one is MolFormer, developed by IBM Research, which leverages Masked Language Modeling and uses a linear attention Transformer combined with rotary embeddings \cite{Ross2022}. It was pretrained on a combined set of 1.1 billion SMILES strings from ZINC and PubChem and learns a compressed representation of chemical molecules. During pretraining, molecules were canonicalized using RDKit to ensure consistency while those longer than 202 tokens were excluded to maintain model efficiency. The pretrained MolFormer model was fine-tuned on various downstream tasks and showed significant improvements in molecular property prediction. It was evaluated on benchmarks such as MoleculeNet, where it showed competitive performance across multiple tasks, including solubility, toxicity, and bioactivity predictions. 

We used both models to encode our set of 1.7 million molecules from ChEMBL and 15 proteins from AlphaFold. The ESM2 model generates an embedding matrix $\mathbf{E} \in \mathbb{R}^{n \times 1280}$, where $n$ is the length of the protein sequence. Given the variability in protein sequence lengths, we pad all embeddings to match the length of the longest protein, which is 1990, resulting in a final matrix $\mathbf{E}_{\text{padded}} \in \mathbb{R}^{1990 \times 1280}$. The MolFormer model produces a fixed-size vector $\mathbf{V} \in \mathbb{R}^{768}$ for each molecule. Formally, for a protein sequence $P_i$ of length $n_i$ and a molecule $M$, their embeddings are represented as:
\[
\mathbf{E}(P_i) = \begin{bmatrix}
e_{1,1} & e_{1,2} & \cdots & e_{1,1280} \\
e_{2,1} & e_{2,2} & \cdots & e_{2,1280} \\
\vdots & \vdots & \ddots & \vdots \\
e_{n_i,1} & e_{n_i,2} & \cdots & e_{n_i,1280}
\end{bmatrix} \in \mathbb{R}^{n_i \times 1280} \quad \text{and} \quad \mathbf{V}(M) = \begin{bmatrix}
m_1 \\
m_2 \\
\vdots \\
m_{768}
\end{bmatrix} \in \mathbb{R}^{768}
\]

\subsection{Final hybrid model}

The docking model, implemented using PyTorch \cite{Pytorch}, is designed to predict the interaction between ligands and proteins through a specialized architecture combining separate sub-models for ligands and proteins. The ligand sub-model is a feedforward neural network, starting with an input dimension of 768, matching the size of the MolFormer embedding. It includes two linear layers with a ReLU activation and a dropout layer for regularization. The protein sub-model uses an LSTM (Long Short-Term Memory) network to process sequential data, taking inputs with a dimension of 1280 to match the size of the input embeddings from ESM2\cite{hochreiter1997long}. The output of the LSTM is further processed through a linear layer to produce features that align in size with the ligand sub-model.

The model's forward pass processes the ligand and protein embeddings through their respective sub-models then concatenates these features into a combined vector. This vector is passed through a regression layer that outputs the docking score prediction. The training phase involves calculating the RMSE between predicted and actual scores and optimizing this loss using the Adam optimizer with a learning rate of 1×10 power -4 \cite{kingma2014adam}. We trained our models on Stanford Carina HPC cluster using a multi-GPU setup with 8 Nvidia Tesla V100 (256 GB of VRAM in total). We tried 8 different variations of our architecture using combinations of parameters for the protein and ligand sub-models sizes, the hidden layers and the dropout rate. Each model was trained for 2 epochs on the train set and used the validation set to print out the RMSE while training to look for signs of overfitting. Finally, the 8 different models were evaluated on the test set in terms of RMSE and $R^2$ score.

\begin{table}[ht]
    \centering
    \begin{tabular}{lcccccc}
        \toprule
        \textbf{Protein Size} & \textbf{Ligand Size} & \textbf{Hidden Layer} & \textbf{Dropout} & \textbf{R² Score} & \textbf{RMSE} \\
        \midrule
        64 & 64 & 128 & 0.4 & -0.01 & 3.73 \\
        64 & 128 & 64 & 0.3 & 0.35 & 3.00 \\
        128 & 128 & 256 & 0.1 & 0.37 & 2.94 \\
        128 & 256 & 64 & 0.5 & 0.36 & 2.97 \\
        128 & 256 & 128 & 0.4 & 0.36 & 2.97 \\
        256 & 256 & 256 & 0.3 & 0.37 & 2.95 \\
        256 & 512 & 64 & 0.5 & \textbf{0.40} & \textbf{2.89} \\
        256 & 512 & 128 & 0.2 & 0.37 & 2.95 \\
        \bottomrule
    \end{tabular}
    \caption{R² Score and RMSE for the LSTM-FFN protein-ligand model}
    \label{tab:results}
\end{table}

\subsection{Hybrid model results}

On Table 3, larger protein and ligand model sizes (e.g., 256) generally lead to better performance, as evidenced by the highest R² score of 0.40 and the lowest RMSE of 2.89 for the configuration with 256 protein size, 512 ligand size, 64 hidden layer, and 0.5 dropout. The impact of hidden layer size and dropout rate on model performance is less pronounced. While there is some variation, configurations with 128 and 256 hidden layers generally performed better. However, no single dropout rate consistently outperformed others, suggesting optimal dropout might be configuration-dependent. The configuration with the smallest protein and ligand sizes (64) and a 128 hidden layer exhibited the poorest performance, with an R² score of -0.01 and RMSE of 3.73, highlighting the inadequacy of smaller sizes for complex protein-ligand interactions.

\section{Limitations and other applications}

\textbf{P2Rank as a probabilistic framework for binding site prediction:} P2Rank uses an ML-based algorithm to predict the binding sites of each protein along with an associated probability. We used an arbitrary threshold to define what counted as a "valid" binding site. In the case where we had multiple binding sites above our 50\% threshold, we only used the one with the highest probability. This means we potentially have missed certain binding sites on the proteins.

\textbf{Conformational space exploration:} A key aspect of molecular docking algorithms like Autodock Vina is the exploration of the search space. We used an exhaustiveness parameter of 8 and tried 5 different poses, the default values for Vina which are known in other studies for balancing accuracy and computational resource use. As our code already took 45 days in total to run, increasing those further would not have been feasible but it could be beneficial for future studies to do a more thorough search. We also limited ourselves to only one binding site per protein, both for computational resources and also to standardize the prediction task for ML researchers. However, it could be interesting to look at algorithms that can work on multiple binding sites at the same time.

\textbf{De novo design of drug-binding proteins:} By using CHeMBL molecules, our dataset has access to a large number of chemical properties. This can potentially be reused for de novo design of drug-binding proteins. Leveraging these properties allows researchers to identify and select molecules with desired characteristics such as high binding affinity, optimal pharmacokinetics, and specific molecular interactions. This enhances the ability to create tailored protein structures that can effectively interact with these molecules, improving the efficiency of the drug discovery process. By incorporating ChEMBL's extensive dataset, researchers can simulate and predict how these molecules might interact with novel protein designs, significantly advancing the development of new therapeutics with targeted actions.

\section{Conclusion and future work}

We introduce Smiles2Dock, an open large-scale multi-task comprehensive dataset for training and benchmarking ML-based protein-ligand docking algorithms. It uses well-know chemical data sources such as AlphaFold and CHeMBL, a diverse set of biologically relevant compounds on the same scale as modern molecular screening databases and is suitable for most major approaches explore such as CNN, graph and embedding based methods. It is easy to use for ML researchers and can be downloaded using two lines of code and a single library using the Datasets library from HuggingFace. We also introduce a novel Transformer-based architecture that uses ESM2 and Molformer to embed molecules and proteins in latent spaces and predict docking scores.

We hope that both chemists and ML practitioners will interact with this dataset and improve on our benchmark, as well as on already existing ML-based docking approaches. To facilitate this, we plan to soon extend Smiles2Dock by increasing the number of proteins used from AlphaFold and also by building a benchmarking website to allow people to submit their results, facilitating comparison between approaches.

\section*{Acknowledgements}

Some of the computing for this project was performed on the Sherlock cluster. We would like to thank Stanford University and the Stanford Research Computing Center for providing computational resources and support that contributed to these research results. We also would like to thank Bharath Ramsundar for his help and feedback on our initial ideas for docking and the model architecture. The content is solely the responsibility of the authors and does not necessarily represent the official views of the funding agencies; funders had no role in study design, data collection and analysis, decision to publish, or preparation of the manuscript. Manuel A. Rivas is in part supported by National Human Genome Research Institute (NHGRI) under award R01HG010140, and by the National Institutes of Mental Health (NIMH) under award R01MH124244 both of the National Institutes of Health (NIH).

\bibliographystyle{unsrtnat}
\bibliography{neurips_data_2023}

\newpage

\end{document}